\documentstyle[12pt]{article}
\input{epsf}
\newcommand{\be}{\begin{eqnarray}}
\newcommand{\ee}{\end{eqnarray}}
\newcommand{\ba}{\begin{array}}
\newcommand{\ea}{\end{array}}
\begin{document}
%
%
\rightline{RUB-TPII-7/98}
\rightline{hep-ph/9806390}
\vspace{.3cm}
\begin{center}
\begin{large}
{\bf Two--pion light--cone distribution amplitudes from the instanton
vacuum} \\
\end{large}
\vspace{1.4cm}
\bf M.V.\ Polyakov$^{\rm 1, 2, a}$ and C.\ Weiss$^{\rm 2, b}$
\\[0.2cm]
$^{\rm 1}${\em Theory Division of Petersburg Nuclear Physics Institute \\
188350 Gatchina, Leningrad District, Russian Federation}\\
\vspace{0.2cm}
$^{\rm 2}${\em Institut f\"ur Theoretische Physik II \\
Ruhr--Universit\"at Bochum \\
D--44780 Bochum, Germany}
\end{center}
\vspace{1cm}
\begin{abstract}
\noindent
We calculate the two--pion light--cone distribution amplitudes in the
effective low--energy theory based on the instanton vacuum. These
generalized distribution amplitudes describe the soft (non-perturbative)
part of the process $\gamma^\ast\gamma \rightarrow \pi\pi$ in the region
where the c.m.\ energy is much smaller than the photon virtuality. They
can also be used in the analysis of exclusive processes such as
$\gamma^\ast p \rightarrow p + 2\pi , 3\pi$ {\em etc.}
\end{abstract}
\vspace{1.5cm} PACS: 12.38.Lg, 13.60.Fz, 13.60.Le
\\
\vfill
\rule{5cm}{.15mm}
\\
\noindent
{\footnotesize $^{\rm a}$ E-mail: maximp@hadron.tp2.ruhr-uni-bochum.de} \\
{\footnotesize $^{\rm b}$ E-mail: weiss@hadron.tp2.ruhr-uni-bochum.de}
%
%
%
%
\newpage
Hadron production in photon--photon collisions at low invariant masses
has become a subject of great interest recently \cite{Brodsky97}. Under
certain conditions such processes are amenable to a partonic
description, if the virtuality of the photon, $Q^2$, is much larger than
the squared c.m.\ energy, $W^2$. The simplest such process, which has
intensively been studied \cite{RR96}, is $\gamma^\ast \gamma \rightarrow
\pi^0$, which provides a unique framework for studying the pion
light--cone distribution amplitude. Recently, Diehl {\em et al.}\ have
argued that also production of a pair of hadrons, $\gamma^\ast \gamma
\rightarrow h\bar h$, can be described in factorized form \cite{Ter}.
The amplitudes for these processes contain new non-perturbative
functions describing the exclusive fragmentation of a quark--antiquark
pair into two hadrons\footnote{Similar functions have previously been
introduced to describe multi--hadron production at large invariant
masses \cite{BG82}.}. In addition to the usual parton momentum fraction
with respect to the total momentum of the hadronic final state, these
functions depend also on the distribution of longitudinal momentum
between the two hadrons, as well as on the invariant mass of the
produced system, $W^2$.  In particular, the authors of Ref.\cite{Ter}
consider the production of two pions, the soft part of which is
contained in a generalized two--pion distribution amplitude.  In
contrast to the one--pion distribution amplitude, which has been studied
using QCD sum rules and other non-perturbative methods \cite{ChZh84,PP},
the two--pion distribution amplitude is an entirely unknown object.
Quantitative estimates of this function are urgently needed for the
computation of the cross section for such processes.
\par
In this note we calculate the two--pion distribution amplitude at a low
normalization point in the effective low--energy theory based on the
instanton vacuum \cite{DP86}. This approach has recently been used to
study the pion and the photon distribution amplitudes \cite{PP,PPRWG98}.
The pion distribution amplitude was found to be close to the asymptotic
one and consistent with the recent CLEO measurements \cite{CLEO97}.
\par
By crossing symmetry the process 
$\gamma^\ast \gamma \rightarrow h\bar h$ is related to virtual Compton
scattering, $\gamma^\ast h \rightarrow \gamma h$, which can be
factorized into a hard photon--parton scattering amplitude and an
off--forward parton distribution (OFPD), which in addition to the
light--cone momentum fraction, $x$, depends also on the longitudinal
component of the momentum transfer to the hadron, $\xi$. The
off--forward isosinglet quark/antiquark distributions in the nucleon
have been recently been computed in the effective low--energy theory
based on the instanton vacuum \cite{PPPBGW97}. It was found that at 
$x = \pm \xi / 2$ the OFPD exhibits characteristic
discontinuities\footnote{Recently, Radyushkin has discussed the behavior
of the off--forward parton distribution near $x = \pm \xi / 2$ in more
general terms using double distributions \cite{Radyushkin98}.}.
Actually, these would--be discontinuities, which would violate
factorization, are reduced to sharp but continuous crossovers by the
fact that the dynamical quark mass derived from the instanton vacuum is
momentum dependent. Given the intimate connection between the OFPD and
the generalized two--particle distribution amplitudes we expect
characteristic discontinuities also in the latter. We indeed find such
would--be discontinuities in the two--pion amplitude considered
here. However, contrary to virtual Compton scattering, in the
$\gamma^\ast \gamma \rightarrow h\bar h$ case they do not seem to lead
to violations of the factorization assumptions.  \par Here we perform a
rough model estimate of the two--pion distribution amplitude at a low
scale ($\sim 600\,{\rm MeV}$), following the approach of
Refs.\cite{PP,PPRWG98}. Our intention is to discuss qualitative features
of these new objects, such as their dependence on the momentum
distribution between the pions, the invariant mass, and isospin, and to
calculate the convolution with the tree--level hard scattering
amplitude. An account of a more extensive investigation will be
published elsewhere \cite{prep}.
\par
{\em Generalized two pion distribution amplitudes.}
Our aim is to compute the generalized two--pion distribution amplitudes
(GDA's), which are defined as \cite{Ter}
\be
\Phi (z, \zeta, W^2 ) &=&
\frac{1}{4\pi} \int dx^- e^{-\frac{i}{2}z P^+ x^-}
\langle \pi^a(p_1) \, \pi^b(p_2) \, |
\, \bar\psi(x) \; \hat n \, T \; \psi(0)\, |0\rangle
\Bigr|_{x^+=0,\>x_\perp=0} .
\label{definition}
\ee
Here, $n$ is a light--like vector ($n^2 = 0$), which we take as
$n_\mu=(1,0,0,1)$. For any vector, $V$, the ``plus'' light--cone
coordinate is defined as $V^+ \equiv (n\cdot V) = V^0 + V^3$; the
``minus'' component as $V^- = V^0 - V^3$. The outgoing pions have
momenta $p_1, p_2$, and $P \equiv p_1 + p_2$ is the total momentum of
the final state. Finally, $T$ is a flavor matrix ($T = 1$ for the
isosinglet, $T = \tau^3$ for the isovector GDA). For brevity we suppress
the isospin indices $a, b$ on the function $\Phi$.
\par
The generalized distribution amplitude, Eq.~(\ref{definition}), depends
on the following kinematical variables: the quark momentum fraction with
respect to the total momentum of the two--pion state, $z$; the variable
$\zeta \equiv p_1^+ / P^+$ characterizing the distribution of
longitudinal momentum between the two pions, and the invariant mass
(c.m.\ energy), $W^2 = P^2$.
\par
In what follows we shall work in the reference frame where 
$P_\perp = 0$.  In this frame
\be
P^-&=&\frac{W^2}{P^+}, \qquad p_1^- \; = \; \frac{W^2(1-\zeta)}{P^+},
\qquad
p_\perp^2 \; \equiv \; p_{1\perp}^2 \; = \; p_{2\perp}^2 \; = \; 
W^2\zeta(1-\zeta)-m_\pi^2.
\hspace{.5cm}
\ee
From these relations one obtains the following kinematical constraint:
\be
\zeta(1-\zeta)\geq \frac{m_\pi^2}{W^2} .
\label{con}
\ee
\par
We shall compute the GDA, Eq.~(\ref{definition}), using the effective
theory of pions interacting with massive ``constituent'' quarks, which
has been derived from the instanton model of the QCD vacuum
\cite{DP86}. The coupling of the pion field to the quarks, whose form is
restricted by chiral invariance, is described by the action
\be
S_{\rm int}&=& \int d^4 x\; \bar \psi(x) \sqrt{M(\partial^2)}
\; U^{\gamma_5} (x) \;
\sqrt{M(\partial^2)}\psi(x) ,
\label{action}
\ee
where
\be
U^{\gamma_5} (x) &=& e^{i \gamma_5 \tau^a \pi^a (x) / F_\pi} 
\;\; = \;\; 1 \; + \; \frac{i}{F_\pi}\gamma_5\pi^a(x)\tau^a-
\frac{1}{2F_\pi^2}\pi^a(x)\pi^a(x) \;\; + \;\; \ldots .
\ee
Here, $F_\pi = 93 \, {\rm MeV}$ is the weak pion decay constant
\par
The momentum--dependent dynamical quark mass, $M(p)$, plays the role of
an UV regulator. Its form for Euclidean momenta was derived in
Ref.\cite{DP86}.  The momentum dependent mass cuts loop integrals at
momenta of order of the inverse average instanton size, 
$\bar\rho^{-1} \approx 600$~MeV.  One should note that the value of the
mass at zero momentum, $M(0)$, is parametrically small compared to
$\bar\rho^{-1}$; the product $M(0) \bar\rho$ is proportional to the
packing fraction of the instantons in the medium, which is a small
parameter fundamental to this picture.  Numerically, a value 
$M(0) = 345 \, {\rm MeV}$ was obtained in Ref.\cite{DP86}.
\par
\begin{figure}
\setlength{\epsfxsize}{12.7cm}
\setlength{\epsfysize}{7.6cm}
\epsffile{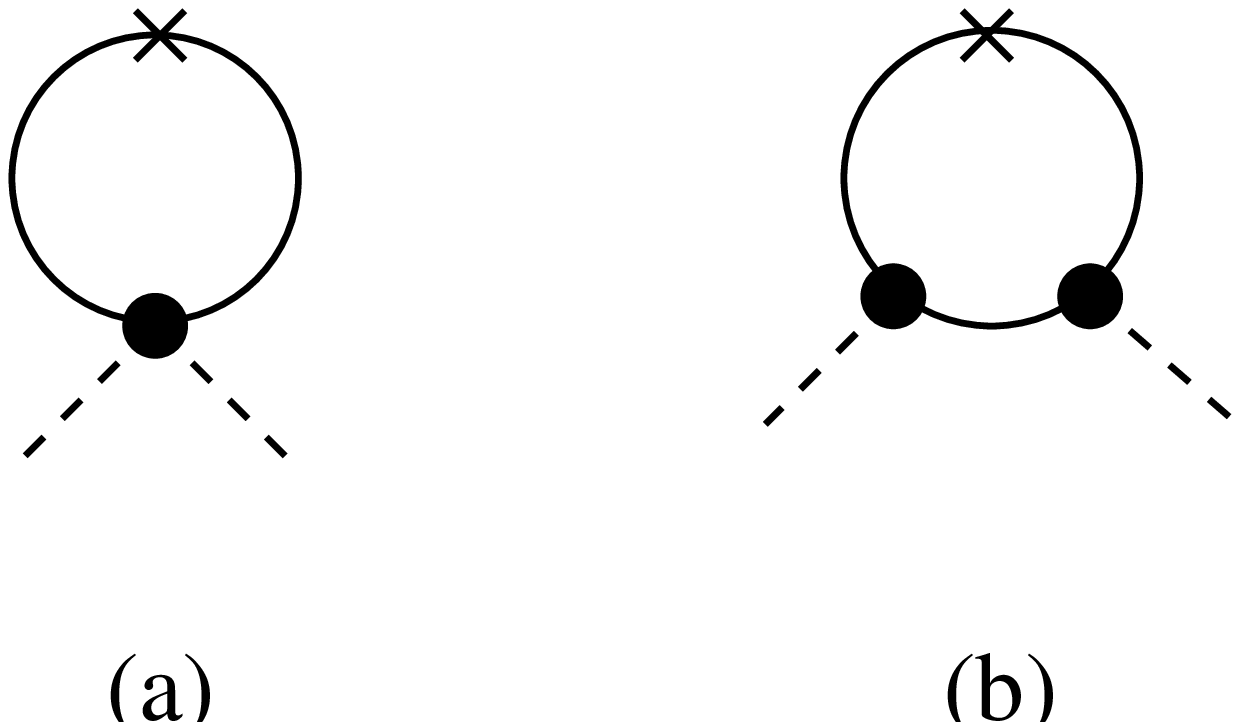}
\caption[]
{The diagrams defining the two contributions to the two--pion
distribution amplitude in the effective low--energy theory, $\Phi^{(1)}$
and $\Phi^{(2)}$.  The solid line denotes the quark propagator with the
dynamical quark mass, the solid points the vertices defined by
Eq.~(\ref{action}).}
\label{fig_fig1}
\end{figure}

{\em Computation of GDA in the effective chiral theory.}
With the simple action Eq.~(\ref{action}) we can proceed to compute the
matrix element defining the GDA, Eq.~(\ref{definition}). In the leading
order of the $1/N_c$--expansion the result for $\Phi(z, \zeta, W^2)$ is
given by the sum of the two Feynman graphs shown in
Fig.~\ref{fig_fig1}. [The quark propagator here contains the dynamical
quark mass.] The first graph, Fig.~\ref{fig_fig1}a, evidently
contributes only to the isosinglet part of the GDA ($T = 1$), whereas
the second graph, Fig.~\ref{fig_fig1}b, contributes to both the
isosinglet and isovector part ($T=\tau^3$).
\par
We first consider the contribution from Fig.~\ref{fig_fig1}a. By
straightforward calculation we obtain
\be
\nonumber
\Phi^{(1)} (z,\zeta,W^2) &=& \frac{4iN_cP^+}{F_\pi^2}
\; \delta^{ab} \; {\rm Tr}(T) \;
\int \frac{d^4 k}{(2\pi)^4} \delta(k^+-zP^+) \sqrt{M(k)M(k-P)}\\
&& \times 
\frac{z \, M(k-P) - (1-z) \, M(k)}{[k^2-M(k)^2+i0][(k-P)^2-M(k-P)^2+i0]}.
\label{Fint}
\ee
From this we immediately see that
$\Phi^{(1)} (z,\zeta,W^2)$ is independent of $\zeta$, and also
that
\be
\Phi^{(1)} (z,W^2) &=& -\Phi^{(1)} (1-z,W^2),
\ee
a property which actually follows from $C$--invariance \cite{Ter}.
\par
\begin{figure}
\setlength{\epsfxsize}{15cm}
\setlength{\epsfysize}{15cm}
\epsffile{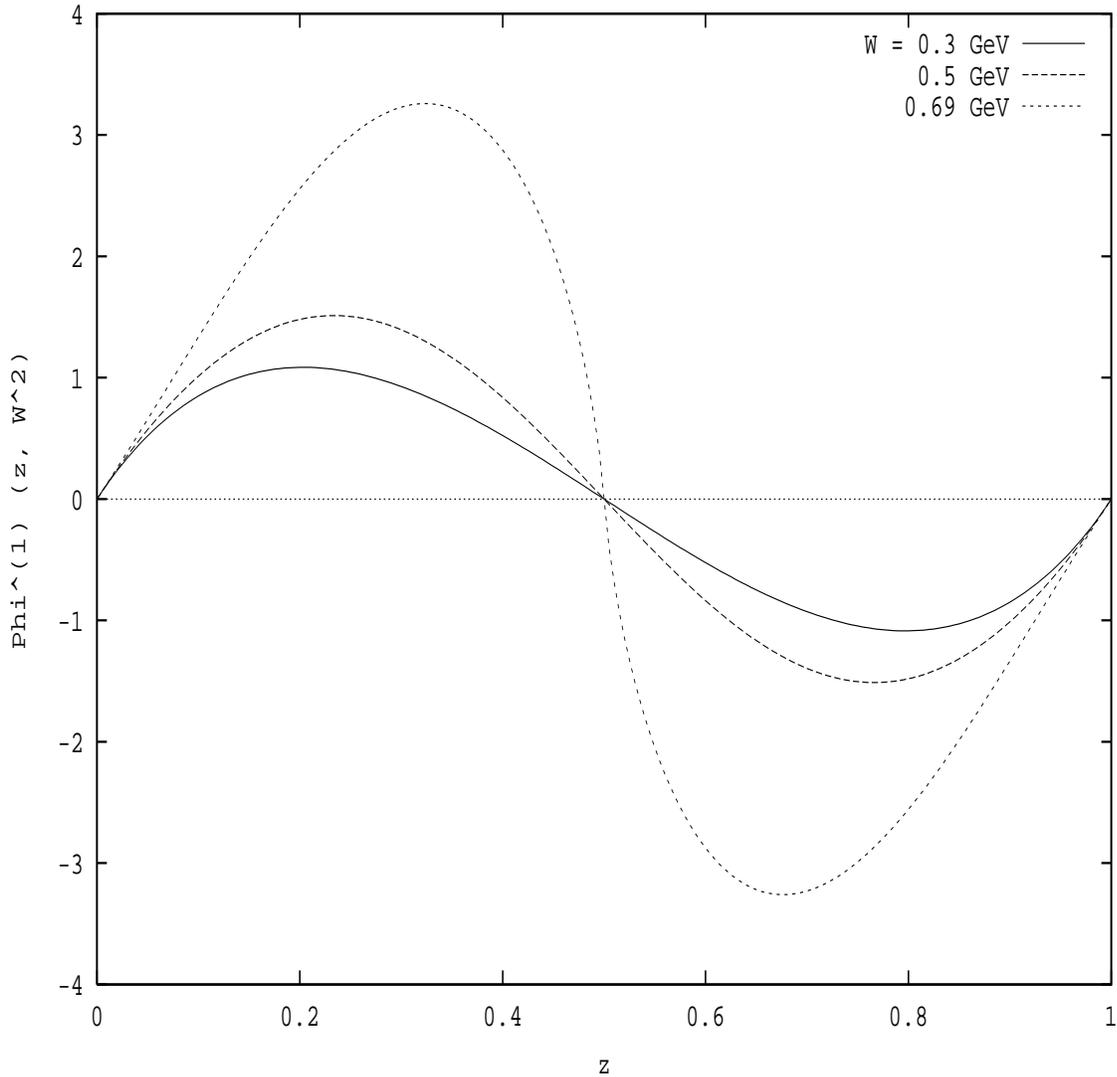}
\caption[]
{The contribution $\Phi^{(1)} (z, W^2)$ to the isosinglet two--pion
distribution amplitude, Eq.~(\ref{Fint}), as a function of $z$ for
various values of $W^2$. This contribution is independent of
$\zeta$. {\it Solid line:} $W = 0.3\, {\rm GeV}$; {\it dashed line:} 
$W = 0.5\, {\rm GeV}$, {\it dotted line:} 
$W = 2 M_0 = 0.69\, {\rm GeV}$.}
\label{fig_fig2}
\end{figure}

Let us first evaluate the integral in Eq.~(\ref{Fint}) neglecting for 
a moment the momentum dependence of the constituent quark mass. In this 
case the result takes a simple form,
\be
\Phi^{(1)} (z,W^2) &=& -\frac{N_c M_0^2}{\pi F_\pi^2}
\; \delta^{ab} \; {\rm Tr}(T) \;
\theta[z(1-z)](2z-1)\int \frac{d^2k_\perp}{(2\pi)^2}
\frac{1}{k_\perp^2+M_0^2-W^2z(1-z)},
\nonumber
\\
\label{Phi_1_const}
\ee
where $M_0 \equiv M(0)$ and $\theta$ denotes the step function.  This
expression is nonzero at the endpoints, $z \rightarrow 0$ and $1$.  Such
behavior of the distribution amplitude would violate the factorization
theorem for the process $\gamma^*\gamma\to\pi\pi$, because the amplitude
for this process contains the integral
\be
I(\zeta,W^2) &=& \int_0^1 dz \frac{2 z-1}{z(1-z)}\Phi(z,\zeta,W^2),
\label{amp}
\ee
which would be divergent if $\Phi$ were nonzero at the endpoints.
However, such conclusion would be premature, as one can easily see that
the momentum dependence of the constituent quark mass becomes crucial at
the end point (see the discussion of this point in
Refs.\cite{PPPBGW97,PP}). This means that when computing the isosinglet
GDA one cannot neglect the momentum dependence of the constituent quark
mass.  For the numerical estimates we can employ a simple numerical fit
to the momentum dependent constituent quark mass obtained from the
instanton model of the QCD vacuum,
\be
M(-p^2) &=& \frac{M_0}{\left( 1 + 0.5 \, p^2 \bar\rho^2 \right)^2}.
\ee
Results of a computation of the distribution are shown in
Fig.~\ref{fig_fig2}, where we have plotted the contribution $\Phi^{(1)}$
to the isosinglet GDA as a function of $z$ at various values of
$W^2$. [Remember that this contribution is $\zeta$--independent].  It is
useful to expand the isosinglet GDA in Gegenbauer polynomials of index
3/2, which are the eigenfunctions of the evolution kernel \cite{ERBL},
\be
\Phi^{(1)} (z,W^2) &=&
\delta^{ab} \, {\rm Tr}(T)\;
6z(1-z) \; \sum_{\scriptstyle n \atop \scriptstyle {\rm odd}}
a_n (W^2) C_n^{3/2}(2z - 1) .
\label{expansion}
\ee
Due to the antisymmetry with respect to $z \rightarrow 1 - z$ only
polynomials of odd degree appear in the expansion of the isosinglet
part. The numerical results for the first few coefficients can in the
momentum range $0\le W^2\le 4 M_0^2$ be approximated by the
forms\footnote{The functions given here should be considered as purely
numerical fits. In particular, the powers in the denominators have no
physical meaning, but are simply chosen such as to represent the
numerical results in the most compact form.}
\be
a_1(W^2)&=&-\frac{0.5556}{\displaystyle 
1 - \frac{W^2}{0.75} }\; ,
\nonumber \\
a_3(W^2)&=&-\frac{0.036(1-6.2 \, W^2)}{\displaystyle 
\left( 1 - \frac{W^2}{0.82}\right)^2}\;,
\nonumber \\
a_5(W^2)&=&-\frac{0.0036}{\displaystyle 
\left( 1-\frac{W^2}{0.5} \right)^{3/2}}\; ,
\ee
where $W^2$ is taken in GeV$^2$. One sees that the coefficients of the
expansion Eq.~(\ref{expansion}) decrease rapidly with increasing the
order of Gegenbauer polynomials. We also give the result for the
contribution of $\Phi^{(1)}$ to the isosinglet part of the integral of
Eq.~(\ref{amp}). With good accuracy the numerical results can be fitted
by
\be
I^{(1)}(W^2) &=& -\delta^{ab}{\rm Tr}(T)
\frac{3.58}{\displaystyle 1 - \frac{W^2}{1.0} }\;.
\label{i1}
\ee
\par
Let us now turn to the second contribution to the GDA, given by the
graph in Fig.~\ref{fig_fig1}b. One can easily convince oneself that the
neglection of the momentum dependence of the constituent quark mass in
this case does not lead to any violation of factorization. Therefore, in
order to make the discussion more transparent we shall in the following
neglect the momentum dependence of the constituent quark mass in the
contribution of graph Fig.~\ref{fig_fig1}b.  The logarithmic divergence
of the loop integrals can always be absorbed in the pion decay constant,
$F_\pi$, which is defined by the logarithmically divergent integral
(written here as an integral over Euclidean momenta)
\be
F_\pi^2 &=& 4N_c \int \frac{d^4 k}{(2\pi)^4} \frac{M_0^2}{(k^2 + M_0^2)^2}.
\ee
All formulae below can easily be generalized to the case of a 
momentum--dependent mass.
\par
Computing the Feynman integral corresponding to graph
Fig.~\ref{fig_fig1}b we obtain for this contribution to the GDA
\be
\nonumber
\Phi^{(2)}(z,\zeta,W^2)&=&\frac{1}{2} {\rm Tr}(T[\tau^a,\tau^b]) \;
\left[\phi^{(2)}(z,\zeta,W^2)-\phi^{(2)}(z,1-\zeta,W^2)\right] \\
&&+ {\rm Tr}(T) \, \delta^{ab} \;
\biggl[\phi^{(2)}(z,\zeta,W^2)+\phi^{(2)}(z,1-\zeta,W^2) \biggr] , \;
\label{Phi_2}
\ee
where
\be
\phi^{(2)}(z,\zeta,W^2) &\stackrel{z< \zeta}{=} &
\frac{ M_0^2N_c}{\pi F_\pi^2}
\int \frac{d^2k_\perp}{(2\pi)^2}\frac{z}{k_\perp^2+M_0^2-W^2z(1-z)}
\nonumber \\
&& \times \left\{ 1 \; + \; \frac{\vec k_\perp\cdot \vec p_\perp+m_\pi^2z-
W^2z(1-\zeta)}{(k_\perp^2+M_0^2)\zeta-2\vec k_\perp\cdot \vec p_\perp z
-m_\pi^2 z+W^2(1-\zeta)z^2}
\right\} ;
\label{fm}
\ee
the value at $z>\zeta$ can be obtained from the above expression as
\be
\phi^{(2)}(z,\zeta,W^2)
&\stackrel{z> \zeta}{=}& -\phi^{(2),(z<\zeta)}(1-z,1-\zeta,W^2).
\label{z_larger_zeta}
\ee
\par
Some comments are in order here. The expressions for the GDA given by
Eqs.~(\ref{Phi_1_const}) and (\ref{Phi_2}, \ref{fm},
\ref{z_larger_zeta}) are valid in the parametrically wide region of
invariant c.m.\ energies $4m_\pi^2 < W^2 < 4M_0^2$.  For $W^2 > 4M_0^2$
the graphs of Fig.~\ref{fig_fig1} develop an imaginary part which is
related to the presence of a two-quark threshold in this model. This
singularity is not physical, because for momenta of order 
$\bar\rho^{-1} \sim 600$~MeV degrees of freedom not accounted for in the
effective theory ({\em e.g.} heavy resonances) start to play significant
role.  These contributions can in principle be estimated
phenomenologically by adding the coupling of the resonances to
constituent quarks to the effective action, Eq.~(\ref{action}). Note
also that the GDA's computed in this model are real at low $W^2$ only in
the leading order of the $1/N_c$--expansion; an imaginary part appears
in the next--to--leading order due to pion loop contributions, which can
be estimated using chiral perturbation theory techniques.
\par
The contribution to the GDA given by Eqs.~(\ref{Phi_2}, \ref{fm},
\ref{z_larger_zeta}) exhibits discontinuities at $z = \zeta$ and 
$z = 1 - \zeta$. The nature of these discontinuities is the same as
those found in the non-diagonal parton distributions computed in this
model \cite{PPPBGW97}.  As was shown in Ref.\cite{PPPBGW97}, the inclusion
of the momentum dependence of the constituent quark mass smears this
discontinuity.  However, in contrast to the case of non-diagonal parton
distributions, the discontinuities in the two--pion GDA do not lead to
violation of factorization, so we can be negligent and evaluate
Eqs.~(\ref{Phi_2}, \ref{fm}) with a constant quark mass.
\par
Let us study the result Eq.~(\ref{fm}) in the chiral limit, $m_\pi=0$,
and expand in powers of $W^2$. In this way we obtain an analytic
expression for the GDA,
\be
&&\phi^{(2)}(z,\zeta,W^2)\stackrel{z< \zeta}{=}
z
\Biggl\{
1+\frac{N_c W^2}{8\pi^2F_\pi^2}\frac{z\left[ 2\zeta 
(1-z)-(1-\zeta)\right]}{\zeta}
+\ldots
\Biggr\}\;,
\nonumber \\
&&\phi^{(2)}(z,\zeta,W^2)\stackrel{z> \zeta}{=}
- (1-z)
\Biggl\{
1+\frac{N_c W^2}{8\pi^2F_\pi^2}\frac{(1-z)\left[ 2(1-\zeta)z
-\zeta \right]}{(1-\zeta)}
+\ldots
\Biggr\}\;.
\label{Phi2_analytic}
\ee
From this expression we find for the lowest two moments
\be
\int_0^1 dz \; \Phi^{(2)}(z,\zeta,W^2) &=& \frac12 (2\zeta-1)
{\rm Tr}(T[\tau^a,\tau^b])
\Biggl\{1+\frac{N_c}{24\pi^2F_\pi^2}W^2+\ldots
\label{m1}
\Biggr\}\;, \\
\nonumber
\int_0^1 dz \; (2z-1) \Phi^{(2)}(z,\zeta,W^2) &=&
\delta^{ab}{\rm Tr}(T)
\Biggl\{\frac 13-2\zeta(1-\zeta)\\
&&+ \frac{N_c W^2}{120\pi^2F_\pi^2}(1-5\zeta(1-\zeta))+\ldots
\Biggr\}\;.
\label{m2}
\ee
The first moment, Eq.~(\ref{m1}), corresponds to the expansion of the
pion electromagnetic form factor for small time--like momenta.  The pion
charge radius read off from Eq.~(\ref{m1}) is
\be
\langle r^2\rangle_{\rm e.m.}=
\frac{N_c}{4\pi^2 F_\pi^2} ,
\ee
which coincides with the result obtained previously in Ref.~\cite{DP86}.
Note that only the isovector part of the GDA contributes to the first
moments, as it should be on grounds of C--parity.
\par
Of $\Phi^{(2)}$ only the isosinglet part contributes to the integral 
Eq.~(\ref{amp}). This contribution to the integral is given by
\be
\nonumber
I^{(2)}(\zeta,W^2) &=& 4\delta^{ab} \, {\rm Tr}(T) \; 
\left\{
-\left( 1+\frac12\log[\zeta(1-\zeta)] \right) \right. \\
&+& \left. \frac{N_c W^2}{48\pi^2F_\pi^2}\left(
4\zeta(1-\zeta)+3\zeta^2\log[\zeta(1-\zeta)]
+3(1-2\zeta)\log (1-\zeta ) \right) \right\} .
\label{i2}
\ee
Note that the R.H.S.\ is symmetric with respect to 
$\zeta \rightarrow 1 - \zeta$.
\par
\begin{figure}
\setlength{\epsfxsize}{15cm}
\setlength{\epsfysize}{15cm}
\epsffile{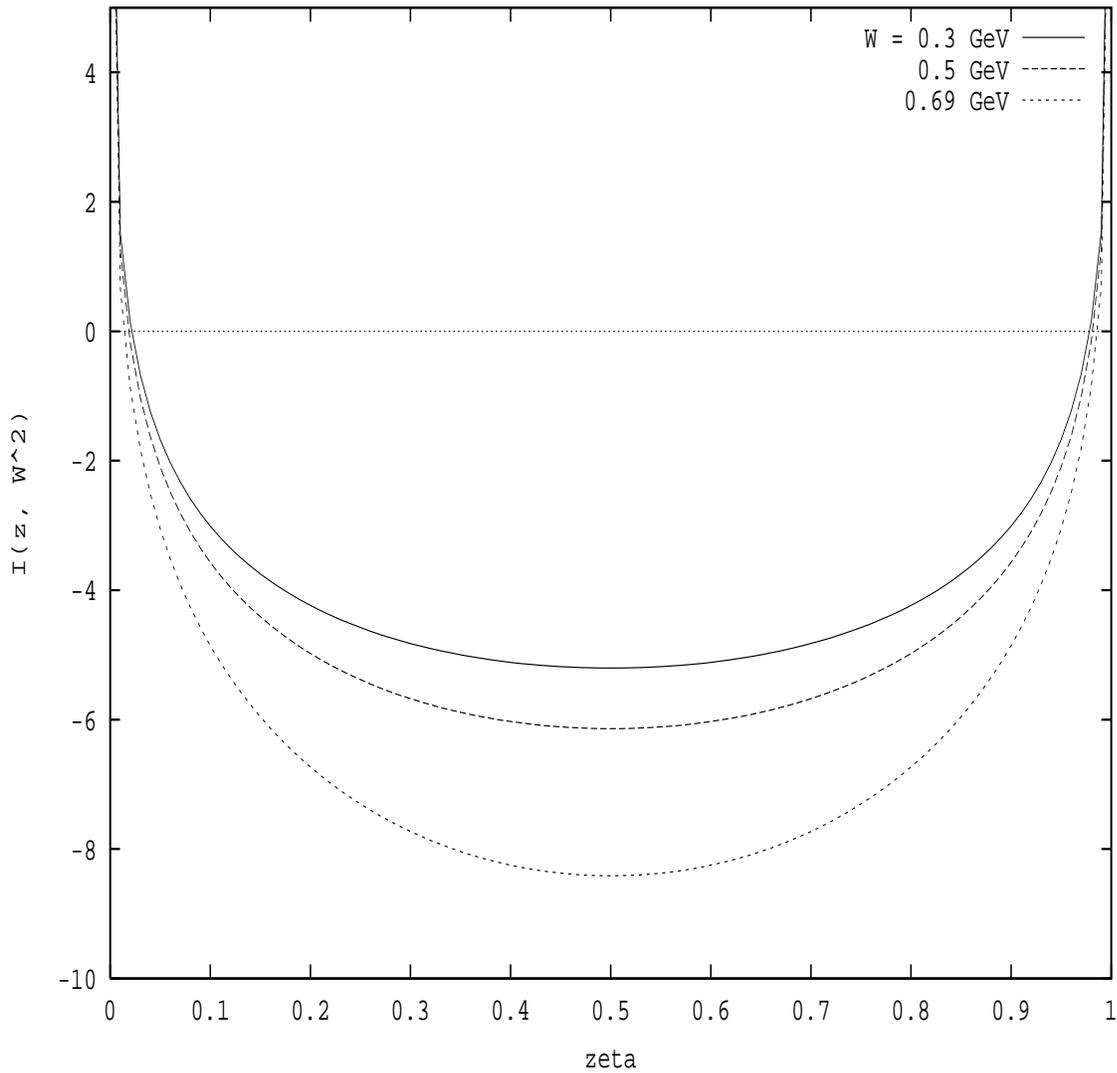}
\caption[]
{The integral determining the $\gamma^\ast\gamma \rightarrow \pi\pi$
amplitude, Eq.~(\ref{amp}), given by the sum of Eqs.~(\ref{i1}) and
(\ref{i2}), as a function of $\zeta$ for various values of $W^2$. {\it
Solid line:} $W = 0.3\, {\rm GeV}$; {\it dashed line:} 
$W = 0.5\, {\rm GeV}$, {\it dotted line:} 
$W = 2 M_0 = 0.69\, {\rm GeV}$.}
\label{fig_fig3}
\end{figure}

The total result for the integral Eq.~(\ref{amp}) is given by the sum of
Eqs.~(\ref{i1}) and (\ref{i2})). In Fig.~\ref{fig_fig3} we plot the
total result as a function of $\zeta$ at several values of $W^2$. We see
that the absolute value of this integral for intermediate values of
$\zeta$ increases with $W^2$, and that the dependence on $\zeta$ is
strong only for small values of $\zeta (1-\zeta)$, a region which is
difficult to access because of the kinematical constraint,
Eq.~(\ref{con}).
\par
\begin{figure}
\setlength{\epsfxsize}{15cm}
\setlength{\epsfysize}{15cm}
\epsffile{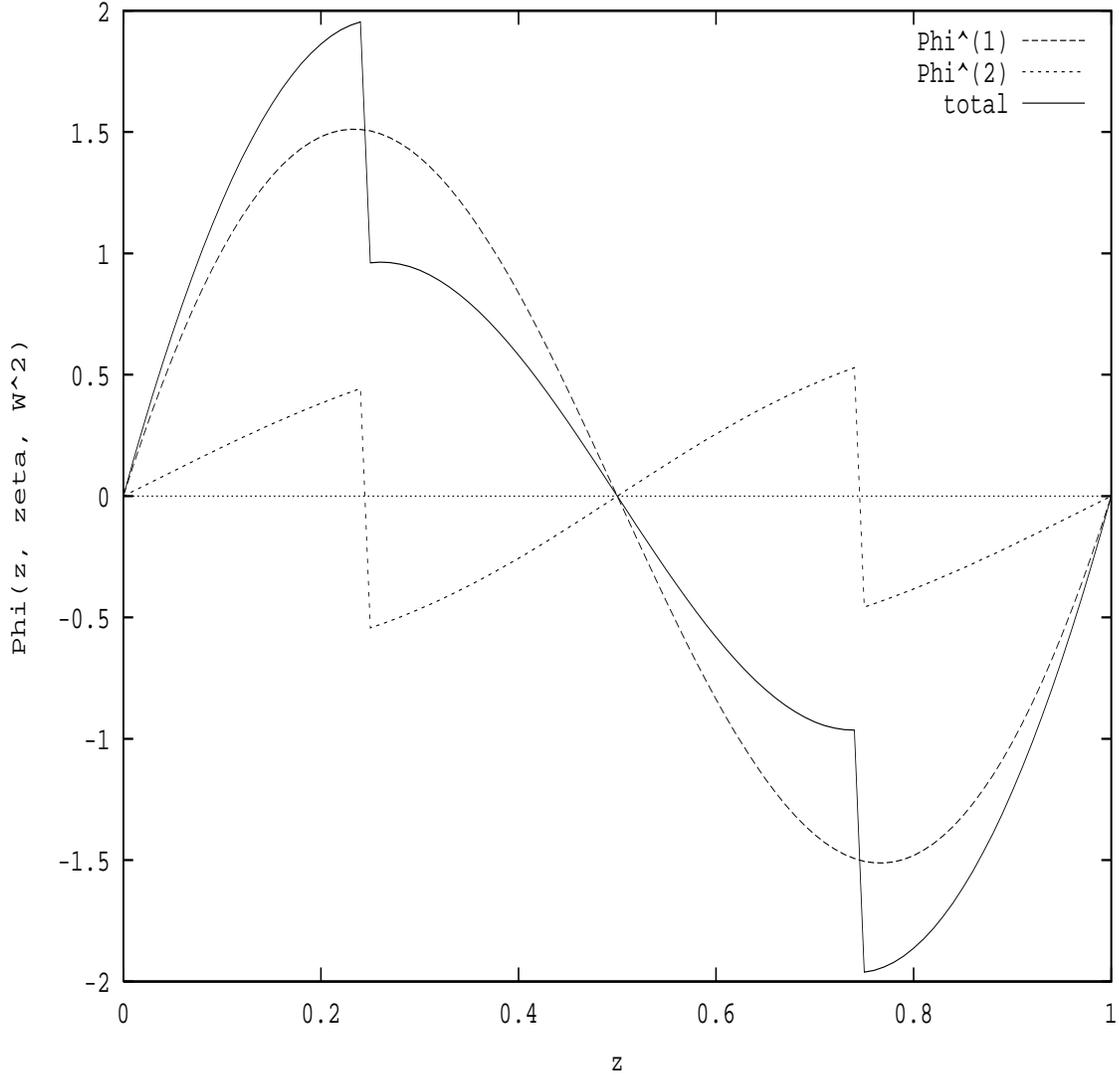}
\caption[]
{The total result for the isosinglet two--pion distribution amplitude,
$\Phi (z, \zeta, W^2)$, as a function of $z$, for $\zeta = 0.25$ and
$W^2 = 0.25\,{\rm GeV}^2$.  {\it Dashed line:} contribution
$\Phi^{(1)}$, Eq.~(\ref{Fint}); {\it dotted line:} contribution
$\Phi^{(2)}$, Eqs.~(\ref{Phi_2}, \ref{fm}, \ref{z_larger_zeta}); 
{\it solid line:} total.}
\label{fig_fig4}
\end{figure}

Finally, in Fig.~\ref{fig_fig4} we show the total result for the
isosinglet GDA, given by the sum of $\Phi^{(1)}$, Eq.~(\ref{Fint}), and
$\Phi^{(2)}$, Eq.~(\ref{Phi2_analytic}), for a value of 
$\zeta = 0.25$. Note the different symmetry of the contributions with
respect to $z \rightarrow 1 - z$, as can be seen also from the analytic
expressions.  We emphasize once more that the discontinuities in
$\Phi^{(2)}$ are smeared out when taking into account the momentum
dependence of the dynamical quark mass.
\par
{\em Asymptotic behavior of the GDA.} 
The scale dependence of the GDA is governed by the usual
Efremov--Radyushkin--Brodsky--Lepage evolution \cite{ERBL,Ter}. This is
obvious when one notes that the hard--scattering kernel of the amplitude
for $\gamma^\ast\gamma\rightarrow \pi\pi$ is the same as that for a
process with only one meson in the final state (with appropriate quantum
numbers). The evolution equations can thus be solved, in analogy to
those for the usual pion wave function \cite{ChZh84}, by expanding in
eigenfunctions of the evolution kernel. In the isosinglet part of the
GDA one has to take into account the mixing with gluon operators
\cite{BG81}; the evolution of the isosinglet GDA calculated in the
effective low--energy theory will be considered elsewhere
\cite{prep}. However, one can easily establish the scale dependence of
the isovector part of the GDA, which does not mix with gluons. This
function does not enter in the amplitude for $\gamma^\ast\gamma
\rightarrow \pi\pi$, but it can appear in exclusive processes of the
type $\gamma^\ast p \rightarrow X p$, where $X$ is a two-- (or more)
pion state, {\em cf.}\ the discussion below.  The asymptotic behavior of
the isovector GDA is given by
\be
\Phi (z, \zeta , W^2 ) &=& 3\, {\rm Tr}\left( T [\tau^a, \tau^b ] \right)
\; z (1 - z) (2 \zeta - 1) F_\pi (W^2 ) ,
\ee
where $F_\pi (W^2 )$ is the pion electromagnetic form factor in the 
time--like domain.
\par
{\em Conclusions.} 
In this note we have given a first model estimate of the two--pion GDA's
at a low scale. A number of qualitative features of the two--particle
distribution amplitudes have emerged, which are consequences of the
general structure of the effective low--energy theory and should thus be
general, even given the crudeness of the quantitative approximations
made.
\par
First, the GDA's generally exhibit discontinuities at $z = \zeta$ and 
$1 - \zeta$, which are in complete analogy to the discontinuities found
in the off--forward parton distribution functions\footnote{In
Ref.\cite{PPPBGW97} only the OFPD's of the nucleon were computed. The
OFPD's in the pion at the low scale exhibit would--be discontinuities
similar to those in the nucleon. A complete study of the OFPD's in the
pion will be published shortly \cite{prep}.} \cite{PPPBGW97}. We have
also noted that, as in the OFPD's, the momentum dependence of the
dynamical quark mass derived from the instanton vacuum turns these
discontinuities into continuous but sharp crossovers. However, since the
discontinuities of the GDA's are not at odds with the factorization of
the amplitude for $\gamma^\ast \gamma \rightarrow h\bar h$ we have been
negligent about implementing the momentum--dependent mass and estimated
the amplitude with the discontinuous GDA's.  Second, we have found that
the isosinglet and --nonsinglet GDA's for the two--pion final state
computed at the low scale show different symmetry properties with
respect to $z \rightarrow 1 - z$.
\par
Our results concerning the $W^2$--dependence of the GDA's and the total
amplitude should be regarded as crude estimates. In particular, the
imaginary part of the functions at $W^2 > 4 M_0^2$ should not be
regarded as physical, since it occurs in a region where the effective
theory is no longer applicable.  A more conservative approach would be
to calculate the GDA's in an expansion in the external pion momenta and
the pion mass, adopting, so to speak, the philosophy of the chiral
Lagrangian.  The physical imaginary part, which is due to final state
interactions, appears in the next order of the $1/N_c$ expansion and can
be estimated using the methods of chiral perturbation theory.  This
approach will be developed in a forthcoming paper \cite{prep}.
\par
We note that the studies reported in this note can easily be generalized
to distribution amplitudes for $n > 2$ pions in the final state. Such
multi--pion GDA's play a role in the analysis of exclusive processes of
the type $\gamma^\ast p \rightarrow X p$, where $X = 2\pi , 3\pi$ {\em
etc.}\footnote{We are grateful to L. Frankfurt and M. Strikman for
discussion of this point.} \cite{BFGMS94}.  For instance, the two--pion
isovector GDA computed here can be used to analyze the $2\pi$ background
in the diffractive leptoproduction of rho mesons.
\par
{\em Acknowledgements.} We acknowledge helpful suggestions by M. Diehl,
as well as interesting discussions with L. Frankfurt, V. Petrov and
M. Strikman. It is a pleasure to thank K. Goeke for encouragement and
support.
%
%

%
%
\newpage
\newpage
\newpage
\newpage
\end{document}